# Fabrication of undoped AlGaAs/GaAs electron quantum dots

Andrew M. See*, Oleh Klochan*, Adam P. Micolich*, Alex R. Hamilton*, Martin. Aagesen[#] and Poul Erik Lindelof[#]
*School of Physics, University of New South Wales, Sydney NSW 2052, Australia
[#]Nano-science center, University of Copenhagen, Universitetsparken 5, DK-2100 Copenhagen, Denmark

*Abstract*—We have fabricated a quantum dot single electron transistor based on an AlGaAs/GaAs heterostructure without any modulation doping. Our device is very stable from an electronic perspective, with clear Coulomb blockade oscillations, and minimal drift in conductance when the device is set to the midpoint of a Coulomb blockade peak and held at constant gate bias. Bias spectroscopy measurements show typical Coulomb 'diamonds' free of any significant charge fluctuation noise. We also observe excited state transport in our device.

## I. Introduction

Semiconductor quantum dots have been studied extensively due to potential applications in single electron detection [1, 2], and as spin qubits for use in quantum computation [3, 4]. Modulation doped AlGaAs/GaAs heterostructures are a widely used basic platform for producing quantum dots due to the ease of fabrication [5]. However, such devices suffer from significant noise due to charging and discharging of the ionized dopants above the heterointerface [6]. Various methods such as bias cooling [7] and depositing an insulator under the gates [6] have been used in an attempt to reduce the charge noise. However a more optimum solution may be to remove the dopants entirely. In this paper, we present the fabrication and electrical characterization of a quantum dot device produced in a nominally undoped AlGaAs/GaAs heterostructure [8].

## II. Fabrication and Measurements

Our device was fabricated in an AlGaAs/GaAs heterostructure (see Fig. 1(a)). Starting from the undoped GaAs buffer and moving upwards, the heterostructure consists of: a 160 nm undoped AlGaAs barrier, a 25 nm GaAs spacer, and a 35 nm $n+$ GaAs cap used as a metallic gate. The heterostructure was wet-etched to define a Hall bar mesa, and NiGeAu ohmic contacts were then produced using a self-aligned process [8]. This process ensures that the gate overlaps the contacts but remains electrically isolated from them. By applying a positive bias, $V_{TG}$, to the gate, a two-dimensional electron gas (2DEG) is electrostatically induced at the AlGaAs/GaAs interface (red dashed line in Fig. 1(a)), and its density, $n$, can be tuned by $V_{TG}$ as shown in Fig. 1(b). Characterization of the heterostructure gave $n = (-1.09 + 3.42\ V_{TG}) \times 10^{11} \text{cm}^{-2}$ and a mobility of ~ 300,000 $\text{cm}^2/\text{Vs}$ at $n$ ~ 1.8 $\times 10^{11}$ $\text{cm}^{-2}$.

A quantum dot with dimensions 0.54 × 0.47 μm was fabricated using electron beam lithography and a $H_2SO_4$ wet etch to form a ~ 45 nm deep trench that divides the cap layer into seven electrically separate gates as shown in Fig. 1(c/d). The central gate is positively biased at $V_{TG}$ to populate the dot and the adjacent 2DEG reservoirs on either side. The remaining six gates form a pair of quantum point contacts (QPCs) at the left and right in Fig. 1(d) for changing the tunnel barrier transparency, and a pair of plunger gates at the center in Fig. 1(d) for tuning the dot's occupancy. Based on its geometry, electron density and a depletion region of 50 nm at the dot walls, we estimate the dot to contain at most 300 electrons. Electrical measurements were performed using both standard ac (Figs. 2 and 5) and dc (Figs. 3 and 4) configurations with the sample mounted in a dilution refrigerator with a base temperature of 40 mK.

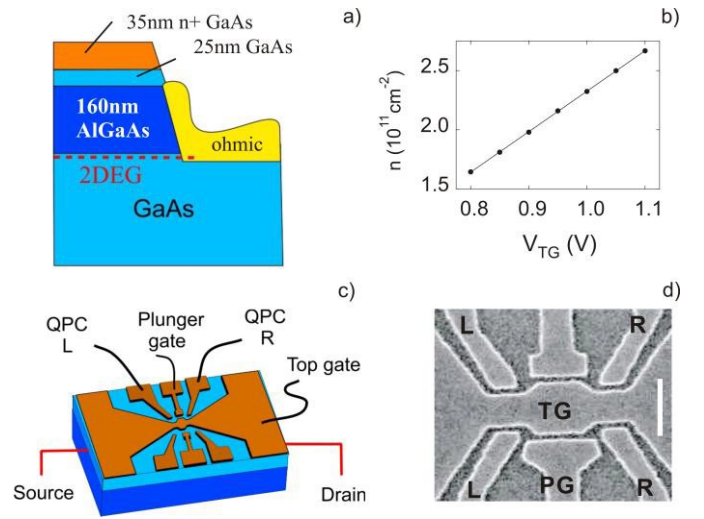

Figure 1. (a) Schematic of the undoped AlGaAs/GaAs heterostructure. (b) The 2DEG electron density, $n$, versus top gate voltage $V_{TG}$. (c) 3D schematic of our quantum dot device showing the top gate etched into seven separate gates with corresponding labels. (d) SEM image of the actual device with the scale bar length of 500 nm.

## III. Results

Fig. 2 shows Coulomb Blockade (CB) oscillations as the plunger gate is swept to change the number of electrons in the dot. The peaks are clean and sharp, with the conductance dropping to zero for extended stretches in between the peaks, suggesting that our dot is the weakly coupled regime [5]. In order to determine the appropriate voltages to apply to the QPC gates to optimize the observed CB oscillations, we performed crosstalk measurements on all three pairs of gates (L, R, and the plunger gate, PG). One such measurement is shown in the inset to Fig. 2, with the two terminal conductance $g$ represented by the color axis, plotted against $V_R$ and $V_{PG}$. The data in Fig. 2 corresponds to a slice along the horizontal white dashed line in Fig. 2 (inset). The bright lines in the inset to Fig. 2 correspond to the CB peaks and the dark regions indicate blockade (i.e., zero conductance through the dot). In the ideal case where there is no cross-talk between the plunger gate and the right QPC, we would expect to see vertical lines instead. This crosstalk behavior is unavoidable due to close proximity of the gates but might be minimized with further design effort.

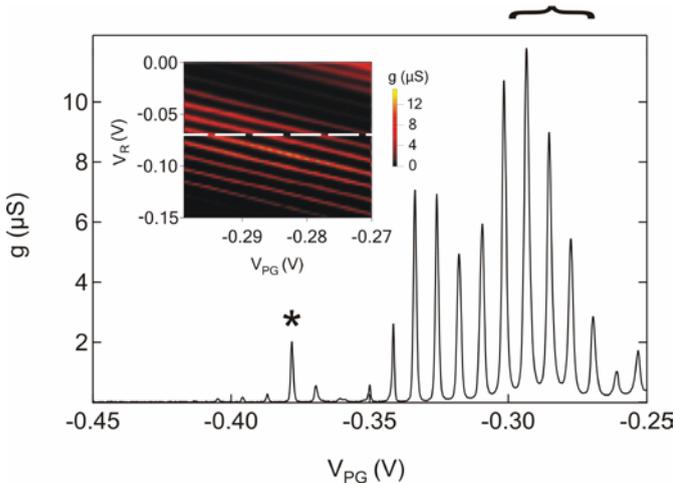

Figure 2. The two-terminal conductance $g$ versus plunger gate voltage $V_{PG}$ at $T$ = 40 mK measured with an ac excitation of $V_{ac}$ = 50 μV. The * indicates the Coulomb Blockade (CB) peak presented in Fig. 3 & 5 where it is indicated by the solid dark and open white circles respectively. (Inset) A colour map of $g$ versus the right QPC bias $V_R$ and $V_{PG}$ obtained with $V_{TG}$ = 0.85 V and $V_L$ = 0 V. The slope of the bright lines away from the vertical indicates crosstalk between the right QPC and the plunger gate. The CB peaks shown in Fig. 2 corresponds to a slice through the inset along the white dashed line; the $V_{PG}$ range of the inset is indicated by a bracket on top of the horizontal axis.

In using quantum dots as ultra-sensitive electrometers [1, 2], it is important to minimize the noise due to charge fluctuations in the device. Thus the basic noise performance of our quantum dot device can be assessed by simply sitting the plunger gate at a fixed voltage, monitoring the current as a function of time, and converting the half peak-peak current noise into an equivalent maximum charge noise. For this study we have chosen the peak indicated by the asterisk in Fig. 2, and this peak is shown in more detail in Fig. 3(a). To maximize the sensitivity in a quantum dot electrometer, we operate the device at a voltage $V_{PG}$ that corresponds to the middle of the rise in the CB peak (see solid circle in Fig. 3(a)) [1, 2]. Here the slope $dI/dV_{PG}$ is greatest, and small fluctuations in the device's charge environment appear strongly in the current. In Fig. 3(b) we show the current measured at a sampling rate of 1Hz over 15 minutes with $V_{PG}$ fixed at −0.3427 V, and observe multispectral noise superimposed on a slowly varying background drift in the current. We use the slope $dI/dV_{PG}$ at the operating point (circle in Fig. 3(a)) to convert the half peak-peak current fluctuation into an effective charge noise, and obtain a value of 0.008 times the electron charge, demonstrating the suitability of our devices for ultra-sensitive electrometry applications.

To further quantify the noise performance of our quantum dot, and to be able to compare directly with the noise performance from other quantum dot devices in the literature, we take the data in Fig. 3(b) and perform a Fourier analysis to extract the power spectral density $S_q(f)$ as a function of frequency $f$, as plotted in Fig. 4. Our data spans the range 0.005 – 0.5 Hz and is has the characteristic appearance of $1/f$ noise. We compare our device's noise performance against literature reports from three devices: an undoped Si single electron transistor (SET) [10], a modulation doped n-GaAs radio-frequency (RF) SET [11], and an undoped p-GaAs quantum dot, which has a very similar design [12].

The reported data for the undoped Si SET covers the frequency range from 0.004 to 10 Hz [10]. Our device has similar performance at low frequencies, but at higher frequencies our device performs considerably better. For example, at 0.1 Hz we obtained a power spectral density of ~ 5 × $10^{-4}$ e/Hz$^{1/2}$, which is ~3 times better than the undoped Si SET (see the star in Fig. 4). Unfortunately the modulation-doped GaAs RF SET was only studied over the range starting from 3 Hz. The modulation doped device gives $S_q$ = 2 × $10^{-4}$ e/Hz$^{1/2}$ at 3 Hz (see circle in Fig. 4), and given that $1/f$ noise becomes more pronounced at lower frequencies, this makes our device considerably quieter in comparison. Finally, the undoped p-GaAs quantum dot has an $S_q$ in the range 2 − 8 × $10^{-4}$ e/Hz$^{1/2}$ [12], giving a comparable noise performance to the undoped n-GaAs quantum dot we present in this paper. Together, these comparisons highlight the superior noise performance that can be obtained with our induced AlGaAs/GaAs single electron transistor technology.

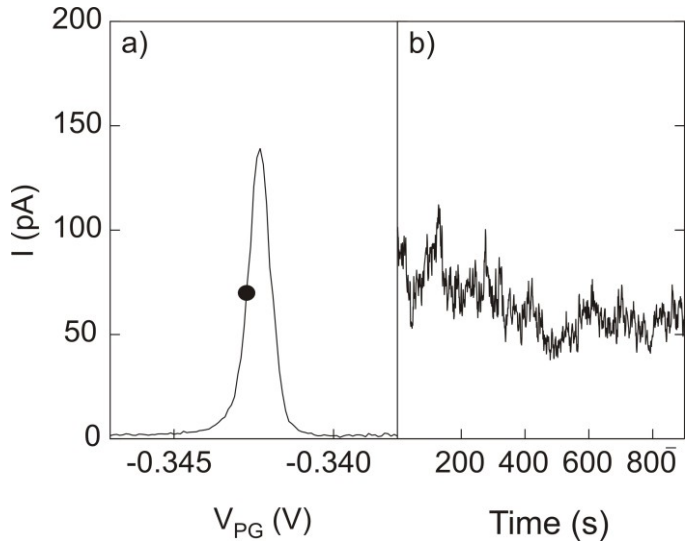

Figure 3. (a) The Coulomb blockade peak indicated by the * in Fig. 2. The solid circle located at $V_{PG}$ = −0.3427 V indicates the point where voltage is held fixed in order to obtain the measurement of current vs time presented in (b). This current consists of a slow background current drift with multispectral noise superimposed thereon, and corresponds to a maximum charge noise of 0.008$e$, where $e$ is the electron charge.

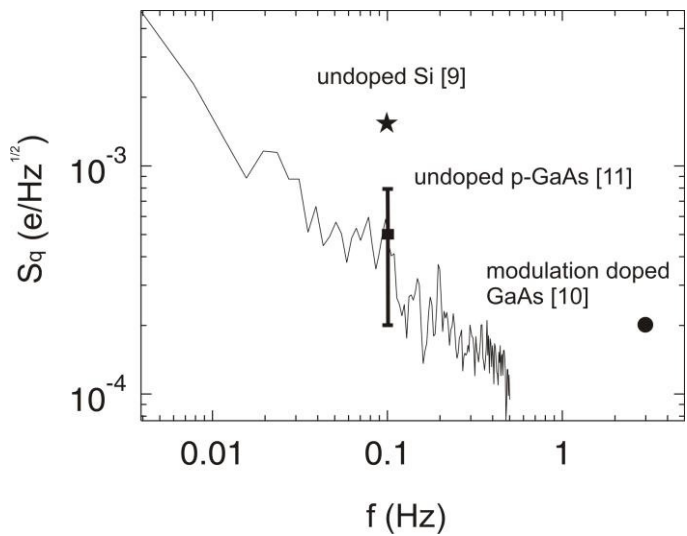

Figure 4. Power spectral density $S_q$ as a function of frequency $f$, obtained from data in Fig. 3(b). Results from three other devices are shown for comparison of the noise performance: the star represents data from an undoped Si SET measured at $f = 0.1$ Hz [9]; the circle represents data from a modulation doped n-GaAs radio-frequency SET measured at 3 Hz [10]; and the square with error bar indicates the range of values obtained from an undoped p-GaAs quantum dot at $f = 0.1$ Hz [11].

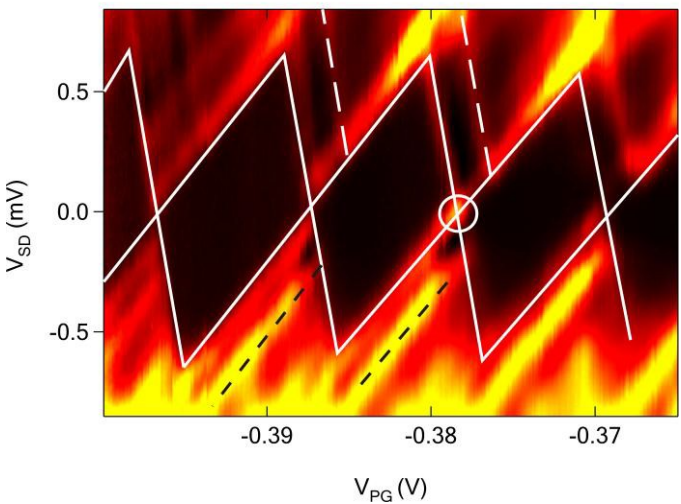

Figure 5. Bias spectroscopy of the quantum dot, showing normalized differential conductance, $g'$ (color axis), against the plunger gate voltage, $V_{PG}$ (x-axis), and dc source-drain, $V_{SD}$ (y-axis). The dark regions correspond to $g'=0$, and form 'Coulomb diamonds' (highlighted by solid white lines). Dashed lines indicate regions where transport via the excited states occurs. The CB peak highlighted by the * in Fig. 2 is located inside the white circle, and corresponds to the peak shown in Fig. 3(a) also.

Additional information is obtained from bias spectroscopy measurements (see Fig. 5), where the differential conductance $g'$ is plotted as the color-axis against the dc source-drain bias $V_{SD}$ and $V_{PG}$. The dark regions indicate low $g'$ and form a sequence of 'Coulomb diamonds' highlighted by the white solid lines. The crossings in between these diamonds correspond to CB peaks, and the dark diamond regions represent blockade with no conductance. The bright lines running parallel to the edges of the diamond (highlighted by the dashed lines), indicate transport via the excited states in the dot [9]. The single particle level spacing of the dot, $\Delta E$, is calculated to be between 180 to 240 µeV [9]. The total charging energy of our dot can be calculated by subtracting $\Delta E$, from the half vertical height of the diamonds, which ranges between 0.44 and 0.45 meV, corresponding to a total dot capacitance of ~ 360 aF. In these Coulomb diamonds, there is no sign of charge noise and random switching events, indicating the stability of our device.

## IV. CONCLUSION

In summary, we have fabricated a quantum dot in an AlGaAs/GaAs heterostructure without modulation doping. We use a heavily-doped cap layer, patterned into gates by electron beam lithography and wet etching to electrostatically control the electron population of the dot. Our device shows clear, stable Coulomb blockade oscillations with transport via excited states in the dot observed in bias spectroscopy measurements. The improved noise performance obtained by removing the modulation doping makes this device architecture interesting for applications such as quantum information and ultra-sensitive electrometry, where very stable quantum dots are extremely useful.